\documentclass[showpacs,a4paper,nofootinbib,tightenlines,floats,twocolumn,superscriptaddress]{revtex4}
\usepackage{times}
\usepackage{bm}
\usepackage{latexsym}
\usepackage{dcolumn}
\usepackage{amsfonts,amssymb}
\usepackage{graphicx,epsfig}
\usepackage{psfrag}
\usepackage{amsmath,amssymb}
\usepackage{graphicx,epsfig}
\usepackage{amssymb}
\usepackage{amsmath}
\usepackage{amsfonts}
\begin{document}
\newcommand {\be}{\begin{equation}}
\newcommand {\ee}{\end{equation}}
\newcommand {\bea}{\begin{array}}
\newcommand {\cl}{\centerline}
\newcommand {\eea}{\end{array}}
\newcommand {\pa}{\partial}
\newcommand {\al}{\alpha}
\newcommand {\de}{\delta}
\newcommand {\ta}{\tau}
\newcommand {\ga}{\gamma}
\newcommand {\ep}{\epsilon}
\newcommand {\si}{\sigma}
\newcommand{\up}{\uparrow}
\newcommand{\down}{\downarrow}

\title{Plaquette valence-bond ordering in $J_1-J_2$  Heisenberg antiferromagnet on the  honeycomb lattice}

\author{H. Mosadeq\footnote{h-mosadeq@ph.iut.ac.ir}}
\affiliation{Department of Physics, Isfahan University of Technology, Isfahan 84156-83111, Iran}

\author{F. Shahbazi\footnote{shahbazi@cc.iut.ac.ir}}
\affiliation{Department of Physics, Isfahan University of Technology, Isfahan 84156-83111, Iran}

\author{S.A. Jafari\footnote{akbar.jafari@gmail.com}}
\affiliation{Department of Physics, Isfahan University of Technology, Isfahan 84156-83111, Iran}
\affiliation{Department of Physics, Sharif University of Technology, Tehran 11155-9161, Iran}
\affiliation{School of Physics, Institute for Research in Fundamental Sciences (IPM), Tehran 19395-5531, Iran}

\date{\today}

\begin{abstract}
We study  $S=1/2$ Heisenberg model on the honeycomb lattice with first and second neighbor
antiferromagnetic exchange ($J_{1}-J_{2}$ model),
employing exact diagonalization in both $S_z=0$ basis and
nearest neighbor singlet valence bond (NNVB) basis. We find that for $0.2<J_2/J_1<0.3$, NNVB basis
gives a proper description of the ground state in comparison with the exact results.
By analysing  the dimer-dimer as well as  plaquette-plaquette correlations and also defining appropriate structure factors, we investigate possible symmetry breaking states as the candidates for the ground state in the frustrated region.
We provide body of evidences in favor of plaquette valence bond ordering for $0.2<J_2/J_1<0.3$.
By further increasing the ratio $J_2/J_1$, this state undergoes a transition to the
staggered  dimerized state.
\end{abstract}


\maketitle
\section{introduction}
Quantum spin liquid (QSL) is  non magnetic  state of a correlated  matter for
which there is no broken symmetry in the spin part of the ground state wave function.
Hence the local magnetic moments  remain disordered down to absolute zero~($T=0$)\cite{balents}.
The quantum ground state for QSL can be expressed as  the superposition of many different configurations,
such as linear combinations of the short range singlet valence bonds. This state is called
resonating valence bond (RVB), originally  proposed by Fazekas and Anderson as
the ground state of the Heisenberg model on the triangular lattice~\cite{RVB}.
The singlet bonds in  RVB state can be considered as pre-formed Cooper pairs, which under suitable conditions~( \textit{i.e.}  hole doping) may coherently propagate throughout the system, hence give rise to superconductivity~\cite{pwan}.

Many  strongly correlated systems are well described by Hubbard Hamiltonian whose ground state for large on site coulomb interaction is the Mott insulating state.
 In this  state  the electrons are localized on the atoms, nevertheless local charge fluctuations induce an Anti-ferromagnetic (AF) exchange interaction between the spins of the electrons. Hence, AF Heisenberg model is an effective Hamiltonian for describing the low energy excitations of
the Mott insulators~\cite{fazekas}. It has been proved that in 1D, the Ground
state of the Heisenberg model is a gapless (critical) spin liquid for $S=1/2$-chain, while it is gapped spin liquid for $S=1$-chain~\cite{Haldane}.
Finding the realizations of QSL in two and three dimensions has been the subject of many researches in  recent years~\cite{balents}. Quantum fluctuations as well as frustration may destroy the long range magnetic order in spin systems. When a spin system is frustrated, it can not find a spin configuration to fully satisfy the interaction between each pair of spins. There are two mechanisms for frustration:
(i) Geometrical frustration, where the lattice geometry is such that, it is not
possible to minimize the interaction energy of all bonds at the same time, \textit{e.g}.
in triangular or Kagom\'e lattice in 2D and pyrochlore lattice in 3D~\cite{Diep}.
(ii) When there are several competing exchange interactions, such as competition
between first and second neighboring AF exchange interactions~($J_1-J_2$ model).
Since the quantum fluctuations are larger in 2D, many attempts to find QSL are focused on the  quasi two-dimensional frustrated spin systems with $S=1/2$~\cite{experiment}.

Two dimensional Heisenberg antiferromagnets apart from their own
importance~\cite{fazekas}, received intensive attention in the
context of layered high-$T_c$ superconducting (HTSC)
material~\cite{pwan}. The ground state of $S=1/2$ Heisenberg model
with AF nearest neighbor (NN) exchange coupling on 2D bipartite
lattices has been shown to be N{\'e}el
ordered~\cite{2D1,2D2,2D3,2D4,2D5,2D6,2D7}. Addition of next
nearest neighbor (NNN) AF interactions frustrates the system and
gradually destroys the (Ne\'el) order. Since AF exchange
interaction encourages the singlet formation, the quantum ground
state of AF Heisenberg model can be expressed in term of over
complete set of valence bond (VB)  basis which represent a total
spin singlet state~\cite{auerbach}.

The $J_{1}-J_{2}$ AF Heisenberg model on square lattice has been extensively studied and various  VB  states have been proposed to
describe its disorder regime~\cite{j1j2square}. One example of such quantum states is nearest neighbor RVB (NNRVB) representing a spin liquid, which  breaks neither translational nor rotational symmetries. However, in highly frustrated regime where the ground state is classically disordered and  the SU($2$) symmetry of the Hamiltonian is restored, there is no theorem to prevent breaking of the lattice translational symmetry. Therefore, in spite of earlier proposal of states with no symmetry breaking~\cite{sq1}, states which break translational symmetry were proposed~\cite{sq2,sq3,sq4}.  One example is
the staggered  dimerized state which breaks translational and
rotational symmetries of the lattice. Another candidate, that has been
proposed recently for some lattices, is plaquette RVB
(PRVB) wave functions in which the resonance of VBs is limited to one
plaquette~\cite{prvb1,prvb3,prvb4,prvb5,prvb6}. PRVB state breaks the translational
symmetry, while preserves the rotational symmetry of the lattice.

\begin{figure}[t]
  \includegraphics[width=5.5cm, angle=-90 ]{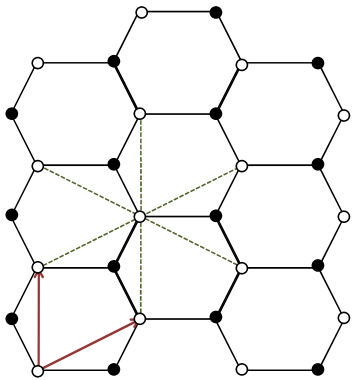}\\
  \caption{The bipartite honeycomb lattice. Two  sublattices are marked by black and white circles. Nearest neighbor lattice points  are connected with solid lines and next to nearest neighbor lattice points are connected with dashed lines. Red arrows show the two primitive lattice vectors.}
   \label{hlattice.fig}
\end{figure}

Recent fabrication of graphene monolayer and also magnetic
compounds with quasi 2D honeycomb structure, has brought the
honeycomb lattice to attention of the physicists from both
experimental and theoretical points of view. Honeycomb lattice does have
coordination number equal to three, which is minimum  among
two-dimensional lattices. In the case of Heisenberg model on
honeycomb lattice, the small number of neighboring interactions
enhances the quantum fluctuations and therefore seems to be
promising system to explore spin liquid states. Honeycomb lattice
is a bipartite lattice composed of two interlacing triangular
sublattices (Fig.~\ref{hlattice.fig}). The unit cell of this
non-bravais lattice contains two sites and lattice is constructed
by two lattice vectors of the triangular bravais lattice. The
non-bravais character of lattice results in more exotic aspects
that can not be seen in square lattice or the other bravais
lattices~\cite{affleck}.

As some realizations of Heisenberg magnets on the honeycomb lattice, one can name
recently discovered  compounds such as InCu$_{2/3}$V$_{1/3}$O$_3$~\cite{exp2} and
Na$_3$Cu$_2$SbO$_6$~\cite{exp1} in which the  Cu$^{+2}$ ions in the copper-oxide layers form
a two-dimensional $S=1/2$ Heisenberg antiferromagnet on a honeycomb lattice,
Bi$_{3}$Mn$_{4}$O$_{12}$(NO$_{3}$) (BMNO) in which the Mn$^{+4}$ ions with $S=3/2$
reside on the lattice points of weakly coupled honeycomb layers~\cite{bmno}.
Replacing Mn$^{+4}$ with V$^{+4}$ in this compound may realize the $S=1/2$
Heisenberg model on honeycomb lattice.
Also the recent progress in the field
of ultracold atoms and trapping techniques~\cite{Aubin} along with
the ability to tune the interaction parameters via the Feshbach
resonance~\cite{Feshbach} can be thought of another way to realize
Heisenberg spins (of localized fermions) on a honeycomb optical lattice.

Two recent achievements has raised the hope of finding QSL in honeycomb geometries.
One, is the large scale quantum Monte Carlo simulation of the half-filled Hubbard
model on the honeycomb lattice, which results  a spin liquid phase
with finite spin gap for moderate values of on-site coulomb interaction ($3.5<U/t<4.3$).
This phase is located between the semi-metallic phase characterized  by massless dirac
fermions ($U/t<3.4$) and the AF-Mott insulating  phase for $U/t>4.3$~\cite{meng}.
The other is the experimentally observed spin liquid behavior in BMNO which
remains magnetically disordered down to $T=0.4$K, in spite of its high
Curi-Weiss temperature $T_{\rm CW}\approx -257 $K~\cite{bmno}.

 Motivated by the above considerations, in this paper we investigate the ground state properties of
 $J_1-J_2$  Heisenberg model which is proposed for explaining the spin liquid behavior of BMNO.
The paper is organized as follows. In section.~\ref{sec2} we introduce  the spin Hamiltonian
and using diagonalization in nearest neighbor
VB basis, we find evidence for spin liquid phase for a range of
coupling constants. In section.~\ref{sec3} we employ
the exact diagonalization in full Hilbert space of $S_z=0$.
With exact wave-functions obtained in this manner, we calculate the
dimer-dimer correlation function. We find two different quantum
phases in frustrate regime. In section.~\ref{sec4} by introducing
suitable structure factors, quantum phase transition point is
determined. At the end, in section.~\ref{sec5}, we calculate
plaquette-plaquette correlations which points to a possible PRVB state.
Section.~\ref{sec6} is devoted to discussion and conclusion.

\section{Model Hamiltonian and its ground state candidates}
\label{sec2}
$J_1-J_2$  AF Heisenberg Hamiltonian  is defined by,
\be
   \mathcal{H}=J_1\sum_{\langle i,j\rangle}S_i.S_j
   +J_2\sum_{\langle\!\langle i,j\rangle\!\rangle}S_i.S_j,
\label{ghamiltoni}
\ee
in which $J_{1}>0$ and $J_{2}>0$ are AF exchange interactions between
first and second neighboring spins, respectively.
The first sum is limited to NN sites, while the second sum
runs over the NNN lattice sites. Since the square lattice
is connected with high $T_c$ superconducting materials,
the studies of frustrated phases of spin models has been
usually limited to this lattice. Recently discovered magnetic materials
with underlying honeycomb geometry is our motivation to study the
above model on the honeycomb lattice.

\begin{figure}[t]
  \includegraphics[width= 5cm, angle=-90]{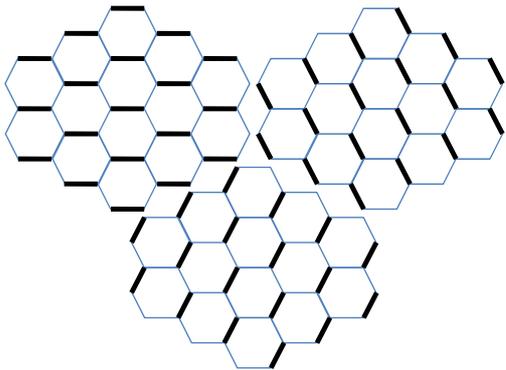}\\
  \caption{Three degeneracy of staggered  dimerized wave function on  honeycomb lattice.}
  \label{cd}
\end{figure}

\begin{figure}[t]
  \includegraphics[width= 5cm, angle=-90 ]{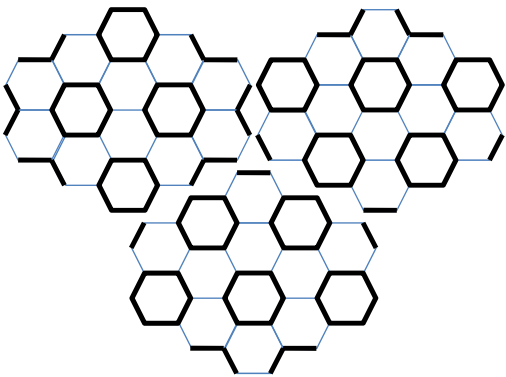}\\
  \caption{Three degeneracy of Plaquette Valence Bond wave function on honeycomb lattice.}
  \label{prvb}
\end{figure}

\begin{figure}[t]
  \includegraphics[width= 5cm, angle=-90]{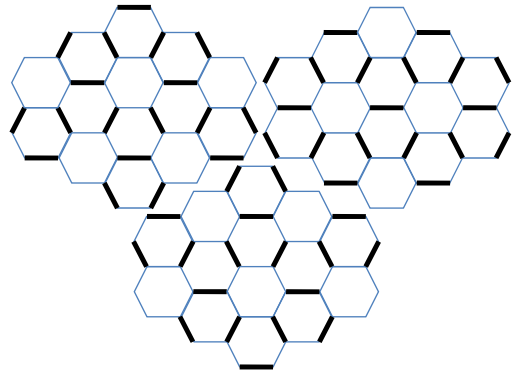}\\
  \caption{Three degeneracy of wave function proposed by Read and Sachdev on honeycomb lattice.}
  \label{rsw}
\end{figure}

Using effective action approach to the frustrated Heisenberg antiferromagnet
in two dimensional system, Einarsson {\em et. al.} proved that the $S=\frac{1}{2}$
disorder ground state on the honeycomb lattice is three-fold degenerate~\cite{einarson}.
Fouet {\em et. al.}~\cite{fouet}  provided some evidence for
staggered  dimerized (SD) state  for $S=1/2$ at $J_2/J_1=0.4$.
Such state breaks the rotational lattice symmetry ($C_3$),
while preserves its translational symmetry (Fig.~\ref{cd}).
They also speculated that spin liquid phase and PRVB phases
can be stabilized for some range of $J_2/J_1$ (Fig~\ref{prvb}).
PRVB phase breaks the translational symmetry, but preserves the
rotational symmetry of lattice. Alternatively,
Read and Sachdev~\cite{rs} used large $N$ expansion method,
proposed another ground state wave function which breaks
both the translational  and the   rotational
symmetries of the lattice (Fig~\ref{rsw}).

In the classical limit (large spins), it has been shown that the
ground state of the above model is N{\'e}el ordered for $J_2/J_1< 1/6$
while for $J_2/J_1> 1/6$ the ground state consists of an infinitely degenerate
set of spiral states characterized by spiral wave vectors $\mathbf q$~\cite{Katsura}.
{ Okumura et al~\cite{Okumura} used a combination of low temperature expansion and
Monte Carlo simulation showed that thermal fluctuations can lift the huge degeneracy 
of the ground state, leading to a state with broken $C_3$ symmetry of honeycomb
lattice. According to their finding, in the vicinity of Ne\'el phase boundary, 
the energy scale associated with the thermal order-by-disorder 
becomes so small that exotic spin liquid behavior, such as 
ring-liquid or pancake-liquid can emerge.}
Mulder {\em et. al.} argued that taking the quantum fluctuations into account,
some specific wave vectors in this manifold are picked as the ground
state -- a manifestation of order by disorder mechanism. They find
for $S=1/2$ quantum fluctuations are strong enough to destroy the spiral
order and stabilize the valence bond solid with staggered  ordering~\cite{Arun}.
Our aim in this paper is to study the ground state of model (\ref{ghamiltoni})
using exact diagonalization in both $S_z$ and
nearest neighbor valence bond (NNVB) basis.

\section{DIAGONALIZATION in NNVB BASIS}
\label{sec3}

The valence bond states are a subset of $S_z$=0 basis with total spin
magnitude ${\bf S}^2$ equal to zero. In this section we show that the ground state
of the $J_1-J_2$ Heisenberg model in
the frustrated regime, where there is no long range order, can be very well
approximated in terms of states in NNVB subspace.

Let us expand the ground state wave function in terms of NNVB states as
\be
   |\psi_0\rangle = \sum_\alpha a(c_\alpha) |c_\alpha\rangle,
   \label{NNVBexp.eqn}
\ee
 where $|c_\alpha\rangle$ denotes all possible configurations $\alpha$ of NNVBs:
\be
   |c_\alpha\rangle=\prod_{(i,j)\in \alpha} (i_\up j_\down-i_\down j_\up).
\ee
First, we have to enumerate the basis $|c_\alpha\rangle$
to construct a numerical representation
of the Hamiltonian matrix in this basis. To determine the basis, the exact Pfaffian
representation of the RVB wave function is employed~\cite{emrrvb}. In this method
one expresses the RVB wave function as the Pfaffian of an antisymmetric matrix
whose dimension is equal to the number of lattice points. The NNVB
basis is much smaller than the whole $S_z=0$ basis, so that the Hamiltonian
matrix can be fully diagonalized with standard library routines.
Note that since the NNVB states ($|c_\alpha\rangle$) are not orthonormal,
one needs to solve the generalized eigen-value problem
$$
   \det [{\cal H}-E{\cal O}] =0,
$$
where ${\cal O}=\langle c_{\beta}|c_{\alpha}\rangle$ denotes the overlap matrix of the NNVB configurations.

\begin{figure}[t]
  \includegraphics[width= 6cm, angle=0 ]{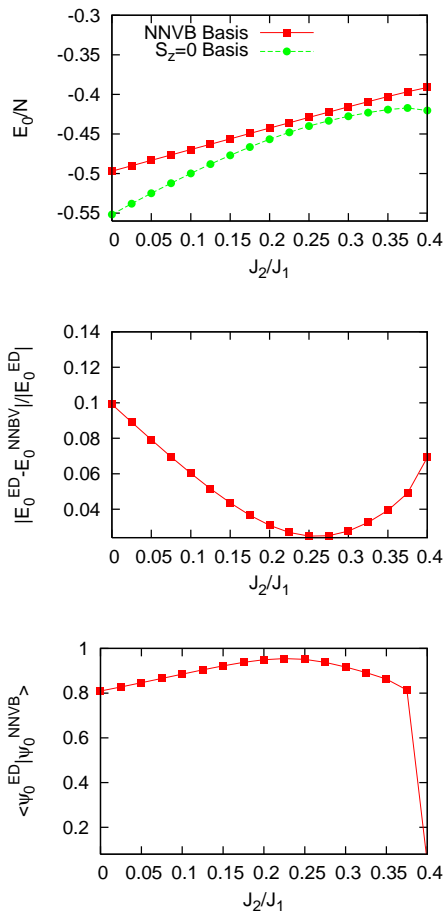}\\
  \caption{Up: The comparison between the exact ground state energy
  evaluated using exact diaganolazition in $S_z$=0 basis (squares) and
  diagonlization in NNVB basis (circles) as a function of $J_2/J_1$.
  middle: The relative errors between the ground state energies obtained by the two basis sets defined as $(E_{0}^{\rm NNVB}-E_{0}^{\rm ED})/({E_{0}^{\rm ED}})$.
Down: The overlap of the exact GS wavefunction $|\psi_{0}^{\rm ED}\rangle$ with the GS wavefunction obtained in NNVB basis $|\psi_{0}^{\rm NNVB}\rangle$.}
\label{esrvb}
\end{figure}

In the upper panel of Fig.~\ref{esrvb} we have compared ground state energies obtained in the NNVB basis,
and those obtained by numerically exact diagonalization in the $S_z=0$ basis versus $J_2/J_1$.
In the middle panel we show the relative error in the ground state energy and the lower panel shows the overlap of the exact ground state wavefunction with the ground state obtained within NNVB basis set.
As can be seen in  this figure, the agreement between the
two sets of energies for $J_2/J_1 \in~ ]0.2, 0.3[$ is remarkable.
Since the NNVB basis is not complete, the large error obtained by NNVB basis for $J_2/J_1<0.2$ and $J_2/J_1>0.3$
can be attributed to the fact that longer range valence bonds
start to contribute. For $J_2/J_1<0.2$, where there is  Ne\'el order in the ground state, it was shown that long-ranged VB states
have remarkable contribution in the ground state wave function~\cite{Noorbakhsh}. Starting from $J_2/J_1=0$, the Ne\'el order is destroyed by
increasing frustration strength up to $J_2/J_1\approx 0.2$. At this point, the spin-spin correlations
will become short ranged and the nature of the ground state can be accurately
captured by NNVB wave functions. For $J_2/J_1>0.3$, the frustrating second neighbor AF exchange coupling $J_2$  induces singlet formation between next nearest  neighbors (NNNVB's) which compete with NNVB's.
Therefore in this region the NNVB basis is insufficient to capture the true ground state. Furthermore, upon increasing $J_2/J_1$ beyond $0.35$, as will be shown shortly, the nearest neighbor singlets in the VB states (dimers) will start to become correlated.

\begin{figure}[t]
  \includegraphics[width= 6cm, angle=-00 ]{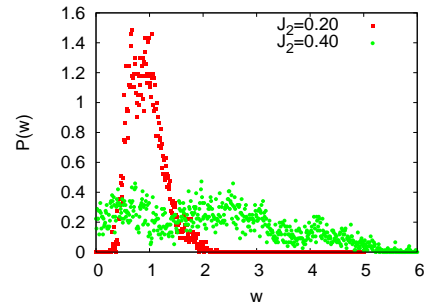}\\
  \caption{Probability distribution  of the relative amplitudes of VB's  in the  ground state wave-function for  $J_2/J_1=0.2, 0.4$  and cluster size  $N=54$.}
  \label{dos}
\end{figure}

\begin{figure}[b]
  \includegraphics[width= 7cm, angle=0]{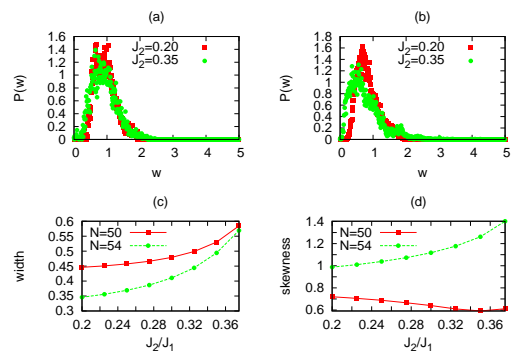}\\
  \caption{ (Color online) PDF of relative amplitude $w$ for $J_2/J_1=0.2,0.35$ for (a) $N=50$ and
  (b) $N=54$. (c) Width and (d) skewness of PDF vs. $J_2/J_1$ for $N=50,54$.
  }
  \label{geometry}
\end{figure}

\begin{table*}[tbhp]
    \centering
        \begin{tabular}{|c|c|c|c|c|c|c|c|c|c|c|c|c|c|c|c|c|c| }
        \hline
    {\bf Trial state}   & \multicolumn{3}{c}{ $\psi_{\rm SD}$} & \multicolumn{6}{|c}{$\psi_{\rm PL}$} & \multicolumn{6}{|c|}{$\psi_{\rm RS}$} \\ \hline
 $\langle P_{\alpha}\rangle_{\rm avg}$     &\multicolumn{3}{c|}{0}  &\multicolumn{6}{c|}{ -0.106(5)} &  \multicolumn{6}{c|}{0} \\ \hline
  {$\alpha'$} & $\beta$ & $\gamma$ & $\delta$ & \multicolumn{2}{c|}{$\beta$} &\multicolumn{2}{c|}{$\gamma$} &\multicolumn{2}{c|}{$\delta$}& \multicolumn{2}{c|}{$\beta$} & \multicolumn{2}{c|}{$\gamma$} & \multicolumn{2}{c|}{$\delta$} \\ \hline
 {$\langle P_\alpha P_{\alpha'} \rangle_1$} & +1/4 & +1/4 & +1/4 & -    & -   & -   & -   & -    & -    & -1/2 &  1   & -1/2&  +1/4& -1/2 &  1   \\
 {$\langle P_\alpha P_{\alpha'} \rangle_2$} & +1/4 & +1/4 & -1/2 & -    & -   & -   & -   & -    & -    & +1/4 & +1/4 & -1/2&  1   & -1/2 & +1/4 \\
 {$\langle P_\alpha P_{\alpha'} \rangle_3$} & +1   & +1   & -1/2 & -    & -   & -   & -   & -    & -    & -1/2 & +1/4 & +1/4& +1/4 & +1/4 & +1/4 \\
  $\langle P_\alpha P_{\alpha'} \rangle_{\rm avg}$&{\bf 1/2}&{\bf 1/2}&{\bf -1/4}&{\bf -}&{\bf -}&{\bf -}&{\bf -}&{\bf -}&{\bf -}&{\bf -1/4}& {\bf +1/2}& {\bf -1/4}&{\bf +1/2}& {\bf -1/4}& {\bf +1/2}  \\ \hline
   $C(\alpha,\alpha')$ & 1/2 & 1/2 & -1/4 & -0.090(5)  & 0.170(8) & -0.090(5) & 0.170(8) & -0.090(5)  & 0.170(8)   & -1/4 & +1/2 & -1/4 & +1/2 & -1/4 & +1/2   \\ \hline
  \end{tabular}
  \caption{$\langle P_\alpha P_{\alpha'} \rangle - \langle P_\alpha \rangle^2$ for $\alpha$
  fixed and $\alpha'=\beta,\gamma,\delta$ (Fig.~\ref{det}). Three indices $1,2,3$ refer to
  three degenerate states (cf. Figs.~\ref{cd},~\ref{prvb},~\ref{rsw}) which become orthogonal to each
  other in the thermodynamic limit for pure SD and RS states. 
  Subscript {\em avg} denotes average over these three possible degeneracies.
  Since the three degenerate PL states are not orthogonal to each other in the
  thermodynamic limit, in this case the correlations have been extracted from 
  finite size scaling of the numeric data. Digits in the parenthesis denote 
  errors in the last digit.
  }
 \label{tab:details}
\end{table*}

 { The RVB wave function (QSL) is defined as linear superposition of all possible VB configurations with the same amplitude. Thus it can be represented as:
\be
   |RVB\rangle = A\sum_\alpha  |c_\alpha\rangle,
   \label{RVB}
\ee
in which $A=\frac{1}{\left(\sum_{\alpha,\alpha'}\langle c_\alpha|c_{\alpha'}\rangle\right)^{1/2}}$ 
is the normalization coefficient.
In order to get preliminary insight into the nature of the ground state obtained in NNVB basis, 
we compare the distribution of the amplitudes  of VB configurations in eq.~(\ref{NNVBexp.eqn}) 
to the uniform amplitude $A$ of the above RVB liquid state. 
For this purpose we define the ratio $w(c_\alpha)=\frac{a(c_\alpha)}{A}$, and look at the
distribution of relative weights, $w$. If the ground state has the characteristics of a RVB 
spin liquid, this distribution is expected to be sharply peaked around  $w=1$. 
In Fig.~\ref{dos} we  plot the  probability distribution function (PDF) of the relative 
amplitudes $w$ for $J_2/J_1 = 0.2$ and $J_2/J_1 = 0.4$ in a cluster of $N=54$ spins. 
As can be seen in this figure, for $J_2/J_1 = 0.2$ the PDF is  narrower  relative 
to $J_2/J_1=0.4$, which implies that the ground state for $J_2/J_1=0.2$ is more similar 
to spin liquid state. The broader distribution of amplitudes for  $J_2/J_1>0.4$ on the other hand, 
indicates significant deviation from QSL behavior, 
which can be considered as a sign of symmetry breaking. To investigate the effect of 
finite lattice geometry on the ground state, we compare PDFs of relative amplitudes 
for two cluster sizes $N=50$ and $N=54$,  and  $J_2/J_1 = 0.2$ and $J_2/J_1 = 0.35$. 
These are shown  in the two top panels of Fig.~\ref{geometry}. 
It can be easily seen that for the lattice size $N=50$, the PDFs are more symmetric than 
$N=54$. This can be attributed to the fact that, in contrast to the cluster with $N=50$ lattice points,
the plaquette wave function can be fitted to the cluster with $N=54$ subject to 
the periodic boundary condition. 
Therefore the symmetry breaking toward plaquette formation is more pronounced for $N=54$. 
This signals a tendency for plaquette formation, provided it is compatible with the lattice geometry.
Moreover for $N=54$, increasing the value of $J_{2}$ leads to emergence of a second peak 
at the right tail of PDF. The average of relative amplitudes of VB configurations taking 
part in the plaquette wave function turns out to be $1.56$. Interestingly, position of the
second peak is quite close to this values. Therefore, the second peak can be attributed  
to the amplification of plaquette character in the ground state, as a result of increasing 
the second neighbor exchange interaction. 
To quantify the variations of PDF ($p(w)$) versus $J_{2}/J_{1}$, we compute its  width given by the standard deviation ($\sigma=\langle (w-\langle w \rangle)^{2}\rangle^{1/2}$)           
 and also its skewness  defined by $\frac{\left\langle (w-\langle w\right\rangle)^{3}\rangle}{\sigma^{^3}} $.  Panel (c) of Fig.~\ref{geometry} is plot  of width versus $J_2/J_1$ for  $N=54$,  showing that the width of PDF  increases monotonically from $J_2/J_1=0.2$ to $J_2/J_1=0.4$. The panel (d) in this figure,  represents the rise  of PDF skewness for $N=54$ in terms of  $J_2/J_1 $, indicating that by increasing $ J_2 $ the distribution functions  for this size get more and more asymmetric, while for $N=50$ skewness does not change remarkably.  
   This suggests that the skewness
can be considered as a heuristic indication of possible symmetry breaking. In summary, 
since the widths of PDFs considered here are finite for the interval 
$0.2\lesssim J_2/J_1\lesssim 0.35$, the above arguments suggest that the ground state of 
$J_1-J_2$ model in this interval is not a perfect QSL state. A more precise understanding 
of the ground state properties, requires the study the correlation between dimers. This will
be done in the following section.}

\section{Exact dimer-dimer correlations}
\label{sec4}

\begin{figure}[b]
  \includegraphics[width= 4cm, angle=-90 ]{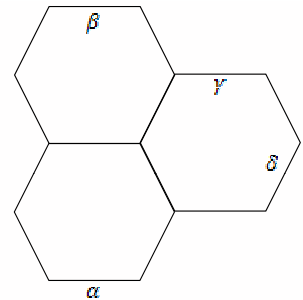}\\
  \caption{The reference bond $\alpha$, and three independent bonds
  $\beta,\gamma,\delta$.}
  \label{det}
\end{figure}

\begin{figure}[t]
  \begin{center}
  \includegraphics[angle=0,width=6cm]{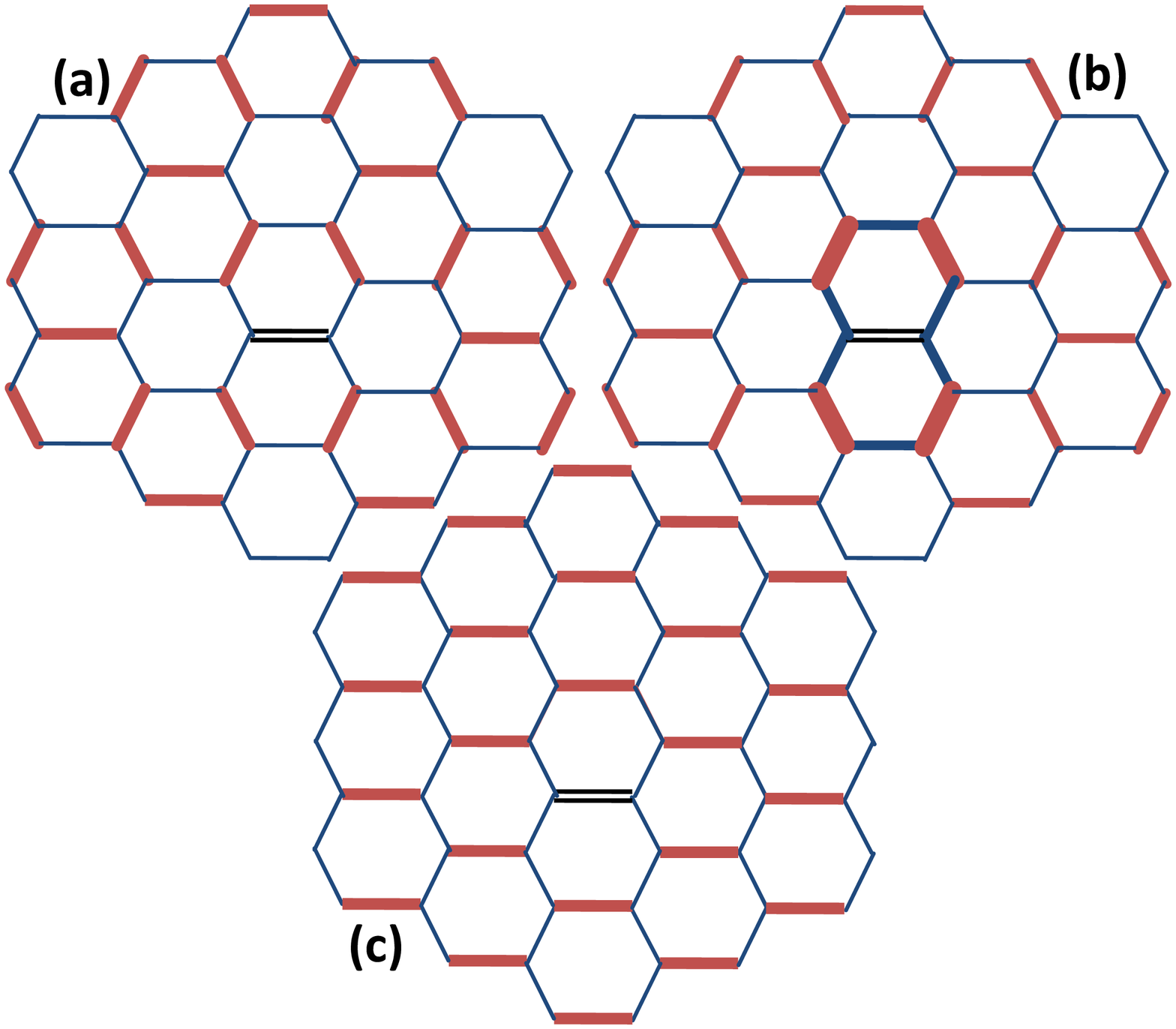}
  \caption{Snapshots of correlations corresponding from left to right
  to (a) RS, (b) PL, and (c) SD trial states. Red and blue links correspond
  to positive and negative correlations with respect to reference bond $\alpha$
  which has been denoted by double line.}
  \label{snapshot.fig}
  \end{center}
\end{figure}

In this section, employing exact diagonalization we obtain the ground state
in $S_z=0$ basis. Using the exact wave function of the ground state,
we calculate correlation between dimers for $0.2\lesssim J_2/J_1\lesssim 0.5$.
The dimer-dimer correlation is defined by,
\be
    C(\alpha,\alpha')=
   4(\langle(\textbf{S}_i.\textbf{S}_j)(\textbf{S}_k.\textbf{S}_l)\rangle-
   {\langle (\textbf{S}_i.\textbf{S}_j)\rangle}^2),
   \label{ddcorr-space.eqn}
\ee
where $\alpha'=(k,l)$, and $\alpha=(i,j)$ is the reference bond relative to which the correlations
are calculated. Define the  permutation operator by,
\be
   P_{kl}=2 ({\mathbf S}_k.{\mathbf S}_l) + \frac{1}{2},
\ee
in terms of which Eq. (\ref{ddcorr-space.eqn}) can be alternatively expressed as
\be
   C(\alpha,\alpha')=\langle P_\alpha P_{\alpha'}\rangle
   -\langle P_\alpha \rangle \langle P_{\alpha'} \rangle.
   \label{calpha.eqn}
\ee
In table~\ref{tab:details} quantities $C(\alpha,\alpha')$ for
fixed $\alpha$ and $\alpha'=\beta,\gamma,\delta$ (Fig.~\ref{det})
are shown for three trial wave functions $\psi_{\rm SD}$, $\psi_{\rm PL}$
and $\psi_{\rm RS}$, where RS, SD and PL  stands for Read-Sachdev,
staggered  dimerized, and plaquette states, respectively.
The expectation values of the operator $\langle P_{\alpha'} \rangle$ for
each of the three degenerate SD and RS states is $-1$ if bond $\alpha'$ is occupied by
a dimer, and $+1/2$, otherwise. 
$\langle P_{\alpha'}\rangle_{\rm avg}$ in table~\ref{tab:details} is
obtained by averaging over three degenerate states corresponding to
SD and RS trial wavefunction.
{  For PL trial state, since the three degenerate configurations are not orthogonal in thermodynamic limit, the averaging is not valid and  we compute $C(\alpha,\alpha')$ numerically 
by finite size scaling method.}   
Fig.~\ref{snapshot.fig} gives a graphical
representation of the correlations obtained  in this way.
Red and blue links denote positive and negative correlations, and
their thickness is proportional to the magnitude of correlations
with the reference bond $\alpha$ of Fig.~\ref{det}.
The reference bond $\alpha$ is denoted by a double line
in Fig.~\ref{snapshot.fig}. { It can be seen in this figure that the pattern for the 
sign of correlations in PL and RS states are identical, except for their magnitudes. 
Specifically, in the PL state, the magnitudes of correlations between the dimers which belong to the
same hexagon as the reference bond, are stronger than the rest of dimers. But in the RS state
all dimers have identical correlation with respect to the reference bond.}

\begin{figure}[t]
  \begin{center}
  \includegraphics[width= 6cm, angle=-90 ]{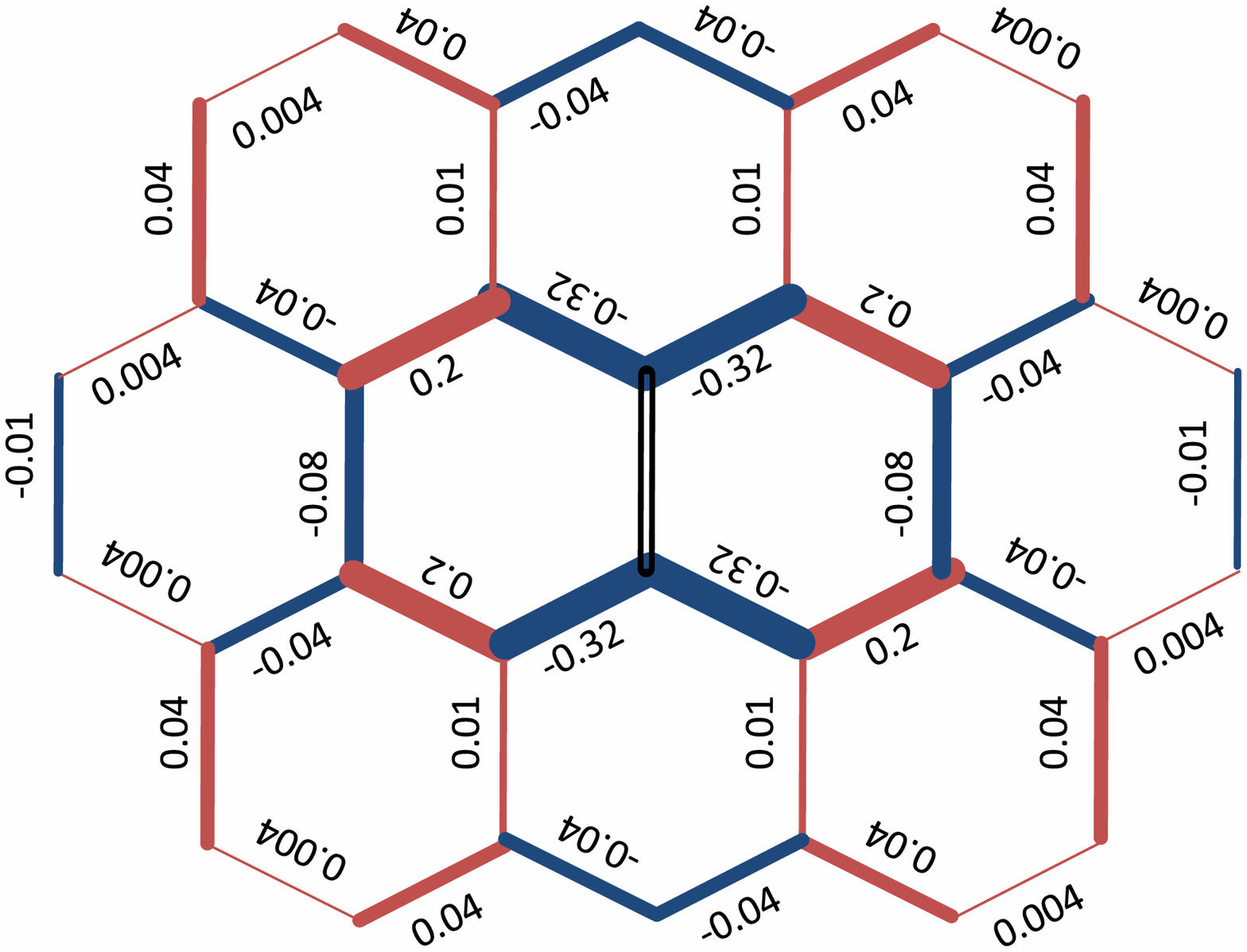}
  \caption{The dimer-dimer correlation for honeycomb lattice with periodic boundary condition
   at $J_2=0.3$. Red (Blue) lines denote positive (negative) correlation.
   The thickness of lines is proportional to the magnitude of correlations.
   The system size is $N=32$.
   }
  \label{dd3}
  \end{center}
\end{figure}

\begin{figure}[t]
  \begin{center}
  \includegraphics[width= 6cm, angle=-90 ]{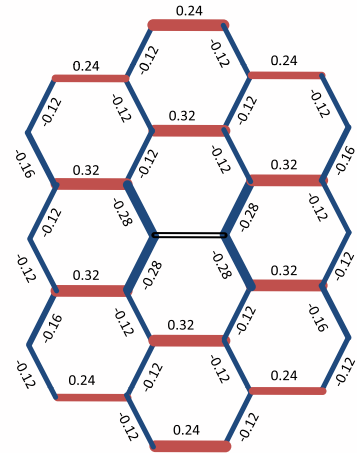}
  \caption{The dimer-dimer correlation for honeycomb lattice with periodic boundary condition
   at $J_2=0.4$. Red (Blue) lines denote positive (negative) correlation.
   The thickness of lines is proportional to the magnitude of correlations.
   The system size is $N=32$.
   }
  \label{dd4}
  \end{center}
\end{figure}

In Fig.~\ref{dd3} and Fig.~\ref{dd4}, we have shown the exact diagonalization results for
the {\em dimer-dimer} correlation functions at $J_2/J_1=0.3$ and $J_2/J_1=0.4$, respectively
for a lattice with $N=32$ sites, subject to periodic boundary conditions.
Note that in order to implement symmetries of infinite lattice on
finite size systems, the size $N$ is limited to specific numbers 
 $N=24,32,42,50,54,\ldots$.
Since the dimension of Hilbert space grows exponentially with $N$,
the exact diagonalization in the whole $S_z=0$ subspace is not feasible,
and hence for $N>32$ we carried out the calculations in NNVB basis.
The correlations are computed with respect to the middle-bond
indicated by double lines. Red bonds denote positive correlations, while
the blue ones indicate negative correlations. The thickness of bonds are
proportional to the magnitude of correlations.
As can be seen in Fig.~\ref{dd3}, for $J_2/J_1=0.3$, the correlations are  decaying
with distance (measured with respect to central bond).
{ Comparing this correlation map with Fig.~\ref{snapshot.fig}(b), shows remarkable
similarity to the snapshot corresponding to PL ordering.}

The dimer-dimer correlation pattern at $J_2/J_1=0.4$ shown in Fig.~\ref{dd4} is entirely
different. First the dimer-dimer correlations do not appreciably decay over the
maximum distance displayed in the figure. Second, one can easily
distinguish a staggered  ordering pattern by noting that, correlations between bonds parallel
to the reference dimer are positive, while others are negatively correlated
with the reference dimer. Comparing with Fig.~\ref{snapshot.fig}~(c),
this correlation snapshot obviously suggests a staggered  dimerized state at $J_2/J_1 = 0.4$.
This is in agreement with previous study of Fouet and coworkers~\cite{fouet}.
However, Fouet {\em et. al.} speculated that at $J_2/J_1=0.3$ the correlation
pattern resembles a RS state. { Based on qualitative symmetry consideration
of short-range dimer-dimer correlations, Fouet and coworkers proposed the possibility 
of crystal of hexagon plaquettes as a candicate for the ground state in the 
$0.3<J_2/J_1<0.35$ range~\cite{fouet}.}
In the following we demonstrate that for
$0.2\lesssim J_2/J_1\lesssim 0.3$, the dominant correlations are of
PL type, rather than RS.

\begin{figure}[t]
  \includegraphics[width= 6cm, angle=0 ]{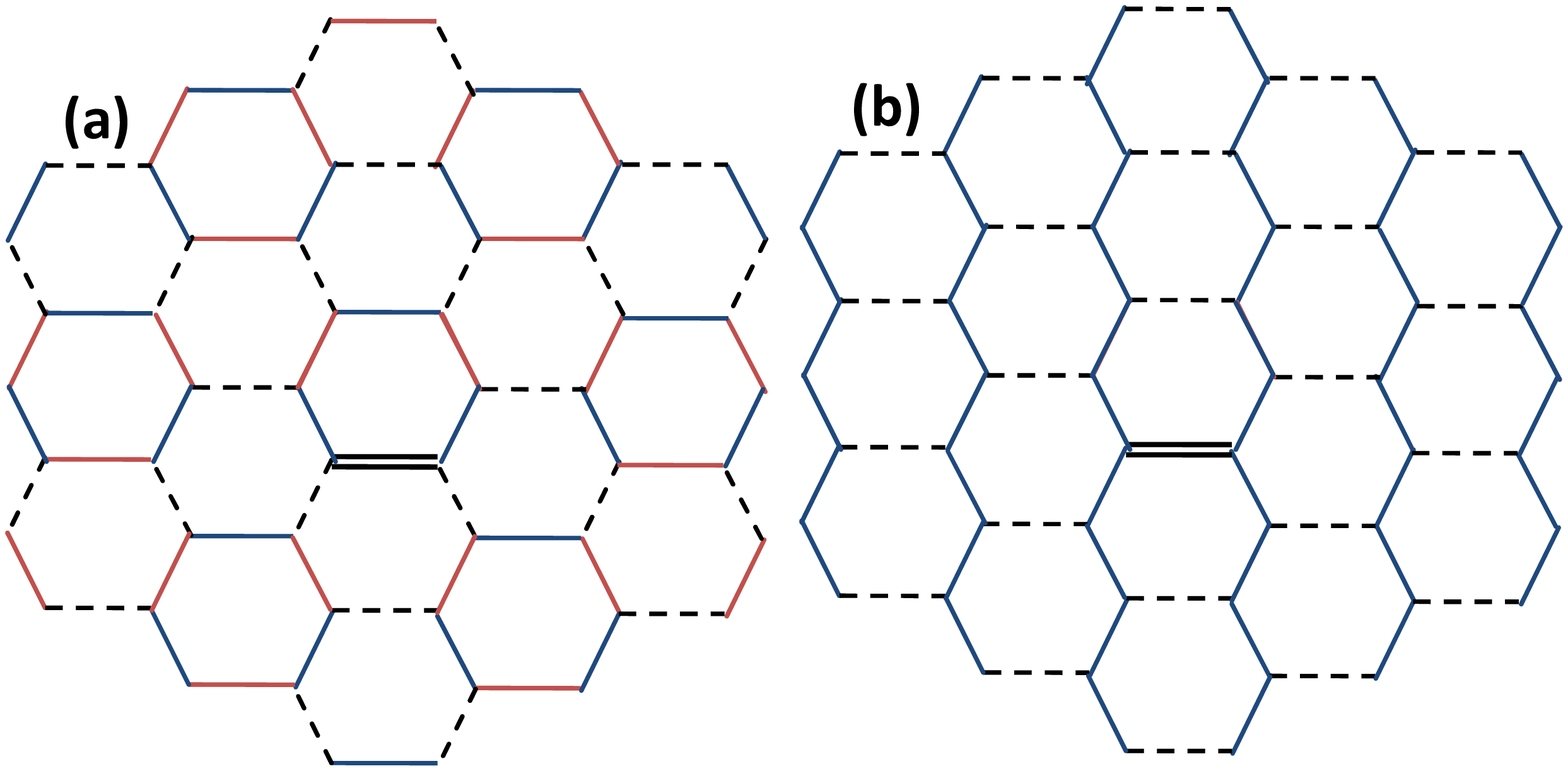}\\
  \caption{ Red and blue links correspond to $+1$, $-1$ phase factors,
  respectively, while the dashed links stand for $0$.
  Reference dimer is identified by double link.
  (a) Phase factor convention for  PL and RS  states.
  (b) Phase factors for staggered  dimerized state.
  }
  \label{phase}
\end{figure}

For quantitative characterization of the nature of VB crystalline state,
we define the following structure factor:
\begin{eqnarray}
  S_\lambda=  \sum_{\alpha'} \varepsilon_\lambda(\alpha')~C(\alpha,\alpha'),
\end{eqnarray}
where $C(\alpha,\alpha')$ is given by Eq.~(\ref{calpha.eqn}) and
$\varepsilon_\lambda(\alpha')$ is the phase factor, appropriately
defined for each of the three states $\lambda\equiv$SD, PL, RS~\cite{prvb5}.
The phase factors $\varepsilon_{\rm SD}, \varepsilon_{\rm PL}$
and $\varepsilon_{\rm RS}$ are shown in Fig.~\ref{phase}.
{ Since the sign of dimer-dimer correlations for PL and RS states are the same, 
their phase factors must be equal.}  
Scaling behavior of $S_\lambda$ for a lattice with
$N$ sites and $N_b$ bonds is given by,
\be
   \frac{S_\lambda}{N_b}=C^\infty_{\lambda} + \frac{A}{N}.
   \label{scaling.eqn}
\ee
Using the above phase factors, and the correlations $C(\alpha,\alpha')$
given in table~\ref{tab:details} we have calculated the corresponding
$C^\infty_\lambda$ in table~\ref{trial.tab} for each of the three trial
states $\psi_\lambda$.

\begin{table}[b]
\caption{Intensive structure factor in thermodynamic limit for the three trial states.}
\label{trial.tab}
\begin{ruledtabular}
\begin{tabular}{c c c c}
Trial state                             & $\psi_{\rm SD}$ & $\psi_{\rm PL}$ & $\psi_{\rm RS}$ \\ \hline
${C^{\infty}}_{\rm SD}$                 &    1/4          &  0              &  0 \\
${C^{\infty}}_{\rm PL}$                 &     0           & 0.125(5)           &  3/8 \\
\end{tabular}
\end{ruledtabular}
\end{table}

\begin{figure}[t]
  \includegraphics[width= 7.5cm, angle=0 ]{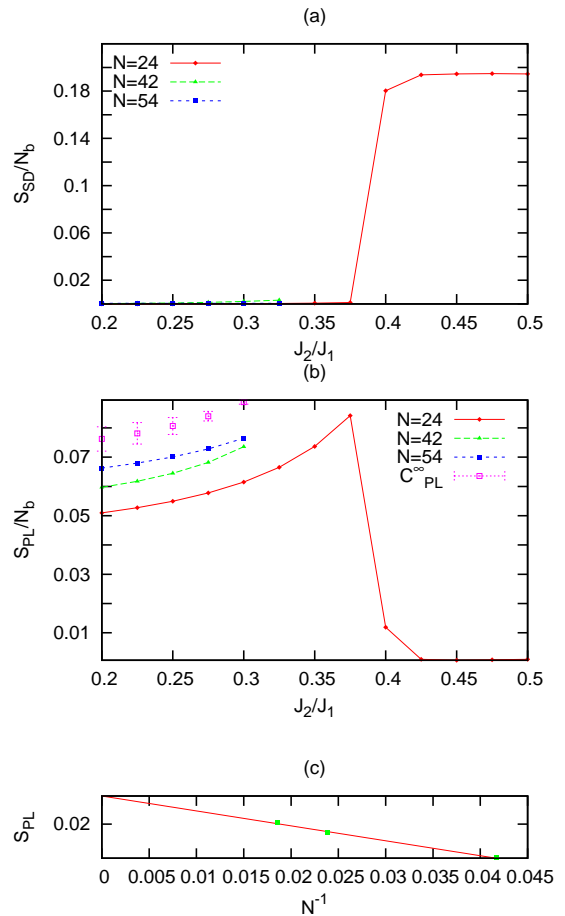}\\
  \caption{The structure factor computed for lattice with $N=24$ (diamond), $N=42$ (triangle)
  and $N=54$ (square). $N_b$ stand for the number of dimers. 
  Structure factors correspond to (a) Staggered  dimerized state, (b) Plaquette state.
  In (b), empty square with error bar indicates extrapolation to infinite lattice size.
  (c) shows a finite size scaling  according to Eq.~(\ref{scaling.eqn}) for PL structure factors
  and $J_2/J_1=0.3$.  
  As can be seen for a specific value of $J_2/J_1$ between $0.35$ and $0.4$, 
  there is a suddent incrase for SD structure factor, which is accompanied by
  a sudden decrease in the PL structure factor.
  }
  \label{sq.fig}
\end{figure}

{ In Fig.~\ref{sq.fig} we have shown $S_{\rm SD}$ and $S_{\rm PL}$ versus $J_2/J_1$
for $N=24,42,54$. Note that for $N=24$, we have done exact diagonalization
in $S_z=0$ basis, while for $N=42,54$, NNVB basis has been used.
Note that the NNVB calculations is valid only in the region $0.2<J_2/J_1<0.35$.
With these three sizes, we have performed finite size scaling according to
Eq.~(\ref{scaling.eqn}) to obtain $C^\infty_{\rm SD}$ and $C^\infty_{\rm PL}$.
Fig.~\ref{sq.fig}(a) and (b) show SD and PL structure factors, respectively, for various sizes. 
For values of $J_2/J_1\gtrsim 0.35$ we do not have reliable data for $N=42, 54$. Hence 
we have not reported finite size scaling for these values of $J_2/J_1$.
For $0.2<J_2/J_1<0.35$, the SD structure factor in (a) remains much smaller than
$C^\infty_{\rm SD}=1/4$ (table~\ref{trial.tab}) for all three sizes.
On the other hand, the PL structure factor in (b) can be extrapolated to a finite value between
$0.07$ and $0.1$ (empty squares) which are comparable with the exact value $0.125$ of
the pure PL state (table~\ref{trial.tab}).

A sudden jump observed in Fig.~\ref{sq.fig} (a) and (b) suggests the existence
of a first order phase transition from PL to SD state as one increases $J_2/J_1$ 
beyond a certain value between $0.35$ and $0.4$.
In panels (b) of Fig.~\ref{sq.fig} the average ratio of structure
factor (averaged over the range $0.2<J_2/J_1<0.35$) to the corresponding
$C^\infty_{\rm PL}$ is given by $(S_{\rm PL}/N_b)/C^\infty_{\rm PL}\approx 0.66$,
while this ratio for RS state in the same region is
$(S_{\rm RS}/N_b)/C^\infty_{\rm RS}\approx 0.08/0.375\approx 0.21$.
Hence for $0.2<J_2/J_1<0.35$ we expect the ground state to be
dominated by plaquette valence bond order. 
}

In addition we calculated the exact value of
$\langle P_{\alpha'}\rangle$ as a function of $J_2/J_1$ for $N=24$ sites.
{ For $J_2/J_1$ from $0.2$ to $0.35$, the expectation value $\langle P_{\alpha'}\rangle$
increases monotonically from $-0.21$ to $-0.12$ and shows a sudden  jumps to $ 0.001$ for 
$J_2/J_1=0.4$}.  In view of the
$\approx -0.1$ value for the expectation value of permutation operator
in the PL state (table~\ref{tab:details}), the negative
values in the range $0.2<J_2/J_1<0.35$ can be considered as
an extra support in favor of plaquette valence bond solid
in this regime.
Guided by the above evidences for plaquette ordering in
region $0.2 < J_2/J_1<0.35$, in the next section we calculate
the plaquette-plaquette correlation using exact diagonalization
method.

\section{PLAQUETTE ORDER in FRUSTRATED REGIME  \label{sec5}}

\begin{figure}
  \includegraphics[width= 6cm, angle=-90]{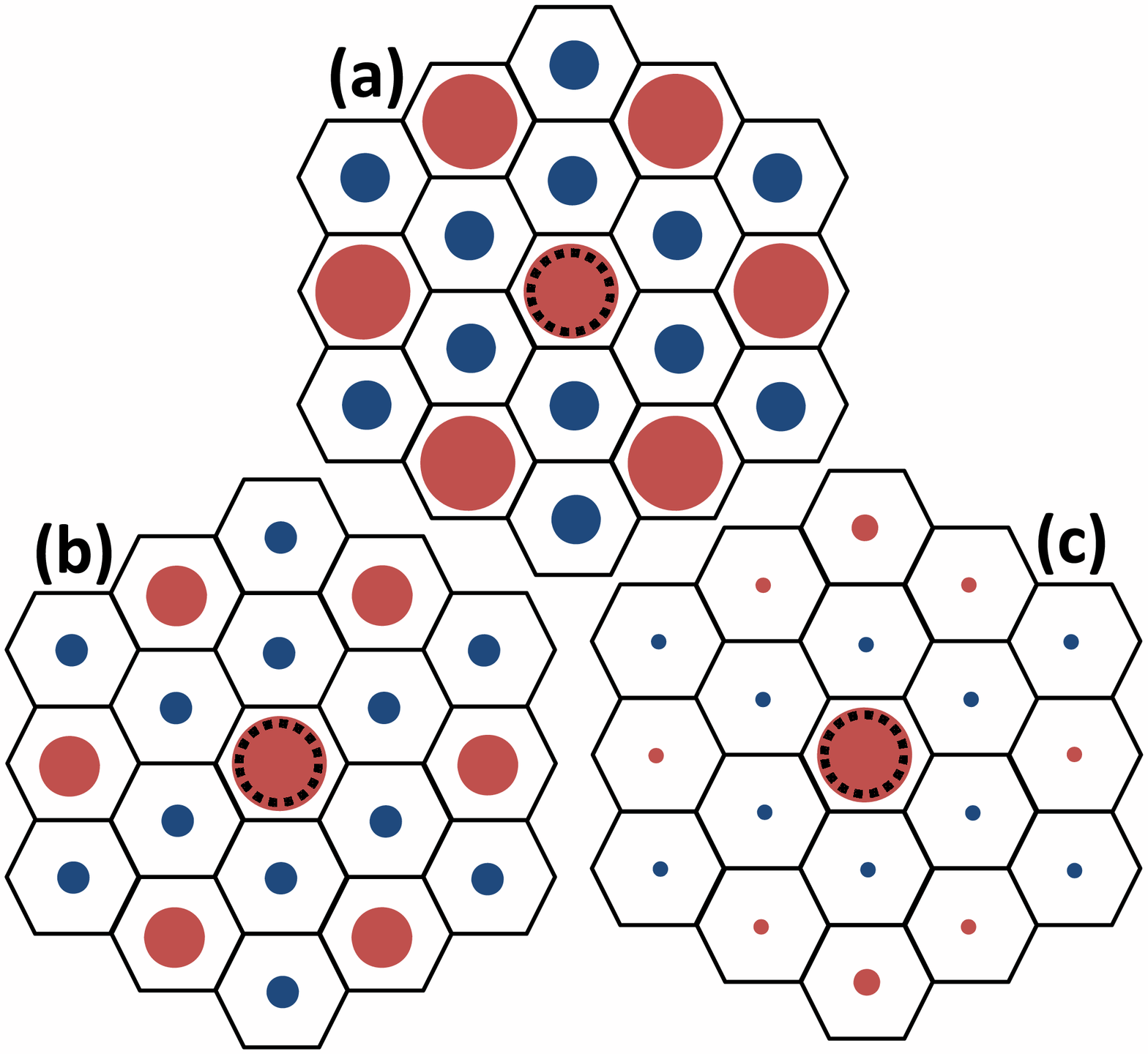}\\
  \includegraphics[width= 6cm, angle=0]{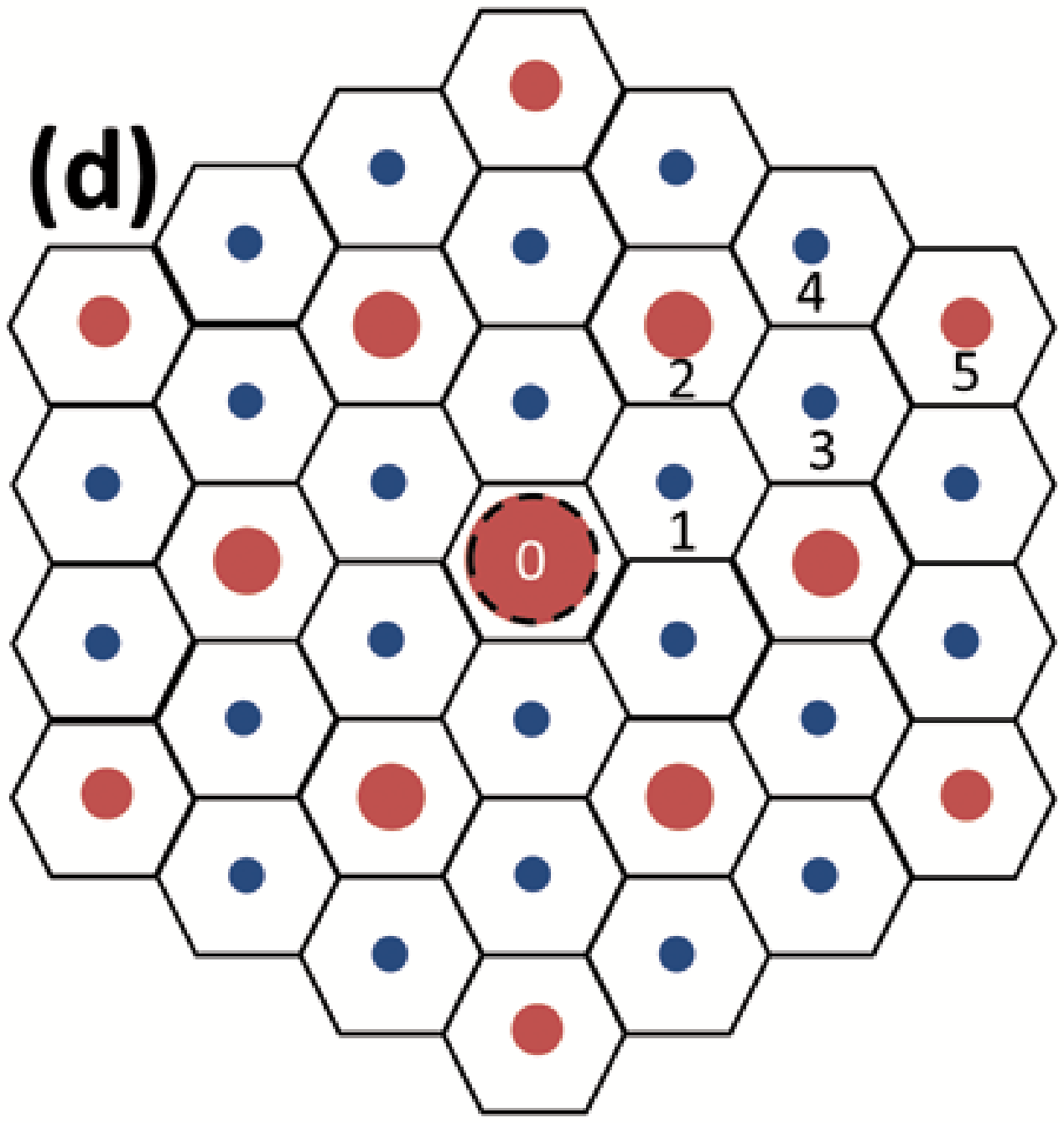}\\
  \caption{  Red (blue) circles denote positive (negative) correlations.
  The radius of circles are proportional to the value of plaquette-plaquette correlation.
  The reference plaquette is depicted by black dashed line.
  (a) The plaquette-plaquette correlation map calculated for PL trial state.
  The exact plaquette-plaquette correlation evaluated using exact diagonalization on honeycomb lattice
  for (b) $J_2/J_1=0.3$ and (c) $J_2/J_1=0.5$. (d) PL-PL correlation pattern for $N=54$ and $J_2/J_1=0.3$.
  }
  \label{ppcor}
\end{figure}

A more direct tool to detect plaquette ordering in
frustrated regime is to investigate the  plaquette-plaquette correlation
defined by,
\begin{eqnarray}
C(p,q)=\langle Q_pQ_q \rangle - {\langle Q_p \rangle}^2,\nonumber\\
Q_p=\frac{1}{2}(\Pi_p+{\Pi^{-1}}_p),
\end{eqnarray}
where $p,q$ stand for different plaquette and $\Pi$ ($\Pi^{-1}$) is
the cyclic exchange operator which permutes six spins around a hexagon
in clockwise (counter-clockwise) direction.
This correlation function was introduced recently and has been used to
investigate plaquette ordering in frustrated Heisenberg magnets
on the checkerboard and square lattice\cite{prvb1, prvb3, prvb4,prvb6}.

In Fig.~\ref{ppcor}-(a) we have depicted the plaquette correlation in
PL state. Red (blue) circles indicate positive (negative) correlations,
with the radii of circles are proportional to the magnitude of correlation.
The reference plaquette is marked with a dashed circle.  Fig.~\ref{ppcor}-(b) and (c) represent  the plaquette correlation
function obtained by  exact diagonalization in $S_z=0$ basis for $N=24$ sites at
$J_2/J_1=0.3, 0.5$, respectively. Comparison of panels (b) and (c) of Fig.~\ref{ppcor}
with panel (a), indicates substantial PL ordering at $J_2/J_1=0.3$.
It is remarkable to note that even the ratio of strengths of positive
and negative correlations in (b) ($\sim 0.29:0.14$) and (a) ($0.35:0.13$)
agree with each other. This ratio in (c) becomes $0.04:0.04$ which significantly deviates from
the corresponding value for a pure PL state (a).
{ Fig.~\ref{ppcor}(d) is the same as (b), for larger size $N=54$ and  $J_2/J_1=0.3$.
As can be seen, for $N=54$ too, a substantial plaquette correlation pattern can be
observed. Moreover, for this size we calculated the PL-PL correlations for  all  hexagons (denoted by 1,2,3,4,5 in Fig.\ref{ppcor}(d))  with respect  to the  reference one to which, number 0 is assigned.  In last column of table \ref{pl.tab} we have listed the relative correlations,	defined by $C(i,0)/C(0,0)$,  obtained by exact diagonalization in NNVB basis and compare them with   corresponding values for PL and RVB states given in first and second columns, respectively. As can be seen, for RVB state this quantity decays rapidly with distance from the reference hexagon, while the exact data does not show such a rapid decaying. This again is an evidence in favor of the PL nature of the ground state. } 

\begin{table}
\caption{Relative plaquette-plaquette correlations for hexagons with different distance to reference hexagon.}
\label{pl.tab}
\begin{ruledtabular}
\begin{tabular}{c c c c}
C(i,0)/C(0,0) & $\psi_{\rm PL}$ & $\psi_{\rm RVB}$ & $\psi_{\rm ED}$ \\ \hline
$i=1   $                 &  -0.12          & -0.04           & -0.08      \\
$i=2   $                 &   0.28          &  0.13           &  0.175      \\
$i=3   $                 &  -0.12          & -0.04           & -0.07      \\
$i=4   $                 &  -0.12          & -0.008          & -0.05      \\
$i=5   $                 &   0.28          &  0.05           &  0.11       \\
\end{tabular}
\end{ruledtabular}
\end{table}

\section{Conclusion  \label{sec6}}
In summary,  diagonalization of $J_1-J_2$ antiferromagnet Heisenberg
Hamiltonian on honeycomb lattice in both $S_z$=0 basis and NNVB basis
show a striking agreement between these two approaches in the parameter range
$0.2 < J_{2}/J_{1} < 0.3$. Therefore, in this region the ground state can be well described in terms of the singlet bonds between the nearest neighbor spins.
Analysis of the exact dimer-dimer correlations,  structure factors, and also plaquette-plaquette correlations, suggests the existence of plaquette valence bond crystal in this range of couplings.
{ In fact the emergence of such PL ordering can be attributed to the 
quantum fluctuations due to the tendency of second neighbors to form singlets.}
This study also reveals a phase transition from the plaquette ordered to the staggered  dimerized state at a point in the interval $J_{2}/J_{1}\in]0.35,0.4]$.
Similar results, regarding  plaquette ordering on the square lattice, have been previously obtained
for $J_{1}-J_{2}-J_{3}$ Heisenberg antiferromagnet in its maximally frustrated region, $J_{2}+J_{3}\sim J_{1}/2$, and for $J_{2}\leq J_{3}$~\cite{prvb5}.
Our results are in contrast with those obtained by QMC simulation of the 
Hubbard model in intermediate interaction regime~\cite{meng}, in a sense that QMC 
results in, a RVB liquid phase with no broken symmetry for this region.
Meng, {\em et. al.} have only calculated short range  dimer-dimer correlation.
Therefore this difference might be due to the fact that the geometry of finite clusters used
in their simulation is not compatible with PL ordering.
Based on present study, we believe that it is necessary to consider lattice geometries
compatible with PL state and to calculate  the plaquette-plaquette
correlation for precise determination of the broken symmetries 
in the ground state.

\section {Acknowledgment}
Numerical analysis in this work have been carried out by \textsf {Spinpack} package~\cite{spinpack}.
This research was supported by the Vice Chancellor for Research
Affairs of the Isfahan University of Technology (IUT). S. A. J was
supported by the National Elite Foundation (NEF) of Iran.
We appreciate useful comments from G. Baskaran, A. Paramekanti, B. Kumar, M. Vojta, A. V. Chubakov.

 \end{document}